\documentclass[showpacs,preprint,pra,superscriptaddress]{revtex4}
\usepackage{amsmath}
\usepackage{graphicx}
\usepackage{dcolumn}
\usepackage{bm}

\begin{document}

\title{}

\title{Quantum phase diffusion of a Bose system: beyond the Hartree-Fock-Bogoliubov approximation}

\author{Ferdinando  de Pasquale}
\affiliation{Dipartimento di Fisica, Universit\`{a} di Roma La
Sapienza, Piazzale A. Moro 2, 00185 Roma, Italy}
\affiliation{CNR-INFM Center for Statistical Mechanics and
Complexity}
\author{Gian Luca Giorgi}\email{gianluca.giorgi@roma1.infn.it}
\affiliation{CNR-INFM Center for Statistical Mechanics and
Complexity} \affiliation{Dipartimento di Fisica, Universit\`{a} di
Roma La Sapienza, Piazzale A. Moro 2, 00185 Roma, Italy}
\pacs{03.75.Hh,03.75.Nt}

\begin{abstract}
A diffusion process is usually assumed for the phase of the order parameter of a Bose system of finite size. The theoretical basis is limited to the so called Bogoliubov approximation. We show that a suitable generalization of the Hartree-Fock-Bogoliubov approach recovers phase diffusion.

\end{abstract}
\maketitle

\section{Introduction}

Bose-Einstein condensation (BEC) of weakly interacting atom systems
attracted a large attention since the studies on liquid Helium, and a
renewed interest stimulated by experimental observations of condensation of
trapped alkali atoms \cite{davis,bradley,anderson}.

Relevant theoretical approaches have been reviewed by Griffin \cite{griffin}. The condensate is usually considered in a coherent state whose amplitude
satisfies a nonlinear Schrodinger equation \cite{ginzburg,pitaevskii,gross}.
Quantum fluctuations determine the instability of this state, known as phase
diffusion (PD), as analyzed in the work of Lewenstein and You \cite
{lewenstein}. This instability is expected in systems in restricted
geometries, where a real symmetry-breaking phase cannot occur. In Ref. \cite
{lewenstein}, a gapless approximation, the Bogoliubov or the Hartree-Bogoliubov (HB), has been shown to exhibit PD, and it is also noted that a complete self-consistent approximation, the Hartree-Fock-Bogoliubov (HFB), has a gapped which spectrum prevents PD. However, it must be rejected because does not satisfy the number conservation law \cite{griffin}.

In a different contest \cite{noi}, we recently described the possibility of
observing BEC and quasi-superfluid behavior in two-mode photon systems. In
that case the single mode approximation can be made. Thus, we obtain the
zero-dimensional version of the problem of two interacting Bose condensates
\cite{castin,javanainen,walls}. This model can be also viewed as a
particular case of \ systems of itinerant polaritons \cite{fazio,plenio}. We
showed that in the study of the evolution of an initial coherent state,
there is, in the condensate phase, an initial time range where the effect of
a symmetry-breaking field can be neglected, and PD should be observed. 

On the other hand, the Bogoliubov transformation amounts to a rotation in
the particle degrees of freedom which leads to quasi-particles. It is easily
shown that the rotation angle becomes infinite in the limit of vanishing
symmetry-breaking field in the HB approximation, while it is kept
finite in the HFB approximation \cite{parkins,dunningham}. Thus, quantities related to the condensate fluctuations, as for instance $\left<aa\right>$ and $\left<a^{\dagger}a^{\dagger}\right>$, in HB are actually diverging in the limit of vanishing
symmetry-breaking field. As a consequence, the instability of the condensate
is strongly enhanced with respect to PD. In spite of the weakness of the
theoretical background, the existence of PD is widely accepted and used to
explain several experimental and theoretical results \cite
{prl98030407,jo,burkov}.

The aim of this work is to show that a self-consistent approximation can be
introduced which satisfies both the requirements of finite rotation and the
absence of a gap in the excitation spectrum. This latter feature is a direct
consequence of the validity of the continuity equation. This result can be achieved within a truncation procedure of the
equations of motion which considers linear and bilinear quantities in the
creation and annihilation operators space. Higher order terms are taken into
account by means of the factorization in the original picture of creation
and annihilation particles operators, or alternatively, by normal ordering
of corresponding quantities in the quasi-particle operators.

The structure of the paper is the following. In Sec. \ref{II} we introduce
the general model and discuss the problem of PD in the Bogoliubov and HFB
approximations. In Sec. \ref{III} we first establish the equations of motion
truncation, and then we show how PD is mainteined in the extended selfconsitent
theory. Then, we conclude the paper in Sec. \ref{IV}.

\section{The model\label{II}}

We first introduce the problem in the context of interacting photon systems \cite{noi}. Two e.m. modes interacts via a tunneling term in a nonlinear medium:
\begin{equation}
H=\omega _{0}\left( a_{1}^{\dagger }a_{1}+a_{2}^{\dagger }a_{2}\right)
-w\left( a_{1}^{\dagger }a_{2}+a_{2}^{\dagger }a_{1}\right) +g\left(
a_{1}^{\dagger }a_{1}+a_{2}^{\dagger }a_{2}\right) ^{2}
\end{equation}
Through a canonical transformation $H$ reduces to
\begin{equation}
H=\left( \omega _{0}-w\right) a^{\dagger }a+\left( \omega _{0}+w\right)
b^{\dagger }b+g\left( a^{\dagger }a+b^{\dagger }b\right) ^{2}
\end{equation}
where $a=\left( a_{1}+a_{2}\right) /\sqrt{2}$ and $b=\left(
a_{1}-a_{2}\right) /\sqrt{2}$. If the value of the hopping constant $w$
exceeds the mode energy $\omega _{0}$, the the vacuum for the mode $a$ is
unstable. Under these conditions, by applying the Bogoliubov approximation
to a coherent state for the mode $a$, a phase diffusion phenomenon appears
\cite{lewenstein}. The interaction between $a$ and $b$ can be seen as a particular case of interaction between condensate and quasi-particles in Bose particle
systems in the limit where only one quasi-particle mode is considered. Taking into account that the ground state of the isolate $b$ system
is the vacuum, we concentrate ourselves on the the parts of $H$ concerning
only the mode $a$: $H_{a}=\omega _{0}^{\prime }a^{\dagger }a+ga^{\dagger
2}a^{2}-\lambda \left( a+a^{\dagger }\right) $, where $\omega _{0}^{\prime
}=\omega _{0}-w+g$, and where we are considering explicitly a
symmetry-breaking field. This
field does not represent a purely mathematical tool introduced to describe the
emergence of a superfluid phase, as usual in boson particles systems, but it can be
physically realized through the interaction of the e.m. modes with a
``classical'' non-fluctuating electron current \cite{glauber}. An exact solution for the evolution of a coherent
state is not known in the presence of the symmetry-breaking field, and different approximations can be made. A constraint
which should be verified by any approach is represented by the continuity
equation
\begin{equation}
\frac{da^{\dagger }a}{dt}=-i\lambda \left( a-a^{\dagger }\right)
\end{equation}

We observe first that in the limit of vanishing nonlinearity the ground
state reduces to a coherent state, that is to the vacuum of a Hamiltonian
where the degrees of freedom have been translated. Then, having in mind a
weak-coupling theory, we perform the translation $a\rightarrow a+\nu $. The
new Hamiltonian is

$H=E_{0}+H_{1}+H_{2}+H_{3}+H_{4}$, where
\begin{eqnarray}
E_{0} &=&\omega _{0}^{\prime }\nu ^{2}+g\nu ^{4}-2\lambda \nu , \\
H_{1} &=&\left( a^{\dagger }+a\right) \left( \omega _{0}^{\prime }\nu +2g\nu
^{3}-\lambda \right) , \\
H_{2} &=&\omega _{0}^{\prime }a^{\dagger }a+g\nu ^{2}\left( 4n_{\alpha
}+a^{\dagger 2}+\alpha ^{2}\right) , \\
H_{3} &=&2g\nu \left( a^{\dagger 2}a+a^{\dagger }a^{2}\right) , \\
H_{4} &=&ga^{\dagger 2}a^{2}.
\end{eqnarray}
In this new representation the continuity equation reads as
\[
\frac{da^{\dagger }a}{dt}+\nu\left( \frac{da}{dt}+\frac{da^{\dagger }}{dt}%
\right) =-i\lambda \left( a-a^{\dagger }\right)
\]

The occurrence of BEC implies $g\nu ^{2}$ finite even for small $g$.

The Bogoliubov approximation amounts to take into account terms of order $%
g\nu ^{2}$ and to neglect terms of order $\sqrt{g}$ and $g$. In this limit
we can disregard \ $H_{3}$ and $H_{4}$. The condensate amplitude $\nu $ can
be fixed by minimizing $E_{0}$:
\begin{equation}
\omega _{0}^{\prime }\nu +2g\nu ^{3}-\lambda =0.  \label{nnumin}
\end{equation}
A finite solution for the condensate amplitude $\nu$, in the limit of vanishing $\lambda$, is obtained only for $\omega _{0}^{\prime }<0$.
We note that this conditions makes $H_{1}$ vanishing. Due to this choice,
\begin{equation}
H=\frac{\lambda }{\nu }a^{\dagger }a+g\nu ^{2}\left( 2a^{\dagger
}a+a^{\dagger 2}+a^{2}\right) .
\end{equation}
For $\lambda /\nu \ll g\nu ^{2}$ \ the ground state is in an eigenstate of
the quadrature $x=\left( a^{\dagger }+a\right) /\sqrt{2}$, while the other
quadrature \ $p=i\left( a^{\dagger }-a\right) /\sqrt{2}$ will have infinite
fluctuations. As far as the evolution of an initial coherent state is
concerned, we obtain PD. Indeed, the coherent state of amplitude $\alpha $
evolves as
\begin{equation}
\left| \alpha \right\rangle _{t}=e^{i\sqrt{2}\alpha p\left( t\right) }\left|
0\right\rangle =\exp \left[ i\sqrt{2}\alpha \left[ p\left( 0\right) -4ig\nu
^{2}tx\left( 0\right) \right] \right] \left| 0\right\rangle .
\end{equation}
The corresponding wave function in the $x$ representation is then
\begin{equation}
\Psi _{\alpha }\left( x,t\right) =\frac{1}{N}\exp \left[ \alpha x-\frac{x^{2}%
}{2}\left( 1-4ig\nu ^{2}t\right) \right] .
\end{equation}
On the other hand, we can characterize PD in terms of the following average quantities
\begin{eqnarray}
\left\langle \alpha \right| x\left( t\right) \left| \alpha \right\rangle
&=&x_{0} \\
\left\langle \alpha \right| p\left( t\right) \left| \alpha \right\rangle
&=&-4g\nu ^{2}x_{0}t \\
\left\langle \alpha \right| p^{2}\left( t\right) \left| \alpha \right\rangle-\left\langle \alpha \right| p\left( t\right) \left| \alpha \right\rangle
&=&\frac{1}{2}+8
g^{2}\nu ^{4}x_{0}^{2}t^{2}
\end{eqnarray}

The HFB approximation amounts to treat $H_{3}$ and $H_{4} $ in a mean-field approximation. If one admits that higher order terms
can correct $H_{1}$ and $H_{2}$ and lead to a squeezed ground state \cite
{parkins,dunningham} which can be defined as the vacuum of $\gamma =\left(
\cosh \theta a-\sinh \theta a^{\dagger }\right) $, $H_{3}$ becomes $2g\nu
\left( 2\sinh ^{2}\theta +\cosh \theta \sinh \theta \right) \left(
a+a^{\dagger }\right) $, while $H_{4}$ is now $g\left[ 4\sinh ^{2}\theta
a^{\dagger }a+\cosh \theta \sinh \theta \left( a^{\dagger 2}+a^{2}\right) %
\right] $. These assumptions being made, the angle $\theta $ is given by the
self-consistent equation
\begin{equation}
\tanh 2\theta =-\frac{2g\left( \nu ^{2}+\sinh \theta \cosh \theta \right) }{%
2g\left( \nu ^{2}-\sinh \theta \cosh \theta \right) +\frac{\lambda }{\nu }}
\label{theta}
\end{equation}
which admits a finite value. While the Bogoliubov approximation satisfies the continuity equation at least in the average, this is not true for HFB, because $dx/dt$ is proportional to $g\sinh \theta \cosh \theta $. 

\section{The extended HFB solution\label{III}}

Now we introduce a mean field approach going beyond the simple Bogoliubov
approximation without introducing a gap in the excitation spectrum. We
discuss the problem starting from the equations of motion. First, we assume
the existence of a finite angle $\theta $ and perform the Bogoliubov
transformation. The Hamiltonian is

\begin{eqnarray}
H_{1} &=&\Lambda \left( \gamma +\gamma ^{\dagger }\right) \\
H_{2} &=&E_{0}\gamma ^{\dagger }\gamma +\frac{1}{2}E_{1}\left( \gamma
^{\dagger }\gamma ^{\dagger }+\gamma \gamma \right) \\
H_{3} &=&2g\nu \left[ \left( \gamma ^{3}+\gamma ^{\dagger 3}\right) \frac{%
e^{\theta }}{2}\sinh 2\theta +\left( \gamma ^{\dagger }\gamma \gamma +\gamma
^{\dagger }\gamma ^{\dagger }\gamma \right) e^{\theta }\left( \cosh 2\theta +%
\frac{\sinh 2\theta }{2}\right) \right] \\
H_{4} &=&g\left[ \left( \cosh ^{2}2\theta +\frac{1}{2}\sinh ^{2}2\theta
\right) \gamma ^{\dagger 2}\gamma ^{2}+\frac{1}{4}\sinh ^{2}2\theta \left(
\gamma ^{\dagger 4}+\gamma ^{4}\right) +\cosh 2\theta \sinh 2\theta \left(
\gamma ^{\dagger }\gamma ^{3}+\gamma ^{\dagger 3}\gamma \right) \right]
\end{eqnarray}
where
\begin{eqnarray}
\Lambda &=&\nu e^{\theta }\left[ \left( \omega _{0}^{\prime }+2g\nu
^{2}\right) -\frac{\lambda }{\nu }+2g\sinh \theta \left( e^{\theta }+\sinh
\theta \right) \right] , \\
E_{0} &=&\Lambda _{0}+g\left[ \cosh ^{2}2\theta +\sinh ^{2}2\theta +\cosh
2\theta \left( 2\sinh ^{2}\theta -1\right) \right] \\
\Lambda _{0} &=&\left( \omega _{0}^{\prime }+4g\nu ^{2}\right) \cosh 2\theta
+2g\nu ^{2}\sinh 2\theta  \nonumber \\
E_{1} &=&\Lambda _{1}+2g\sinh 2\theta \left( \cosh 2\theta +\sinh ^{2}\theta
-\frac{1}{2}\right) \\
\Lambda _{1} &=&\left( \omega _{0}^{\prime }+4g\nu ^{2}\right) \sinh 2\theta
+2g\nu ^{2}\cosh 2\theta
\end{eqnarray}
Then, we build a set equation of motions by limiting ourselves to consider
only linear and bilinear terms and neglecting higher order operators: we
choose the operators $\gamma $, $\gamma ^{\dagger }$, $\gamma ^{2}$, $\gamma
^{\dagger 2}$, and $\gamma ^{\dagger }\gamma $ as independent variables. The
set of coupled equations reads
\begin{equation}
i\frac{d\gamma }{dt}=\Lambda +E_{0}\gamma +E_{1}\gamma ^{\dagger }+3g\nu
e^{\theta }\sinh 2\theta \gamma ^{\dagger 2}+g\nu e^{\theta }\left( 2\cosh
2\theta +\sinh 2\theta \right) \left( \gamma ^{2}+2\gamma ^{\dagger }\gamma
\right)  \label{gamma}
\end{equation}
\begin{equation}
i\frac{d\gamma ^{\dagger }}{dt}=-\Lambda -E_{0}\gamma ^{\dagger
}-E_{1}\gamma -3g\nu e^{\theta }\sinh 2\theta \gamma ^{2}-g\nu e^{\theta
}\left( 2\cosh 2\theta +\sinh 2\theta \right) \left( \gamma ^{\dagger
2}+2\gamma ^{\dagger }\gamma \right)  \label{gamma+}
\end{equation}
\begin{eqnarray}
i\frac{d\gamma ^{2}}{dt} &=&E_{1}+2\left[ \Lambda +g\nu e^{\theta }\left(
2\cosh 2\theta +\sinh 2\theta \right) \right] \gamma +6g\nu e^{\theta }\sinh
2\theta \gamma ^{\dagger }  \nonumber \\
&&+2\left[ E_{0}+g\left( \cosh ^{2}2\theta +\frac{1}{2}\sinh ^{2}2\theta
\right) \right] \gamma ^{2}+3g\sinh ^{2}2\theta \gamma ^{\dagger 2}
\nonumber \\
&&+2\left( E_{1}+3g\cosh 2\theta \sinh 2\theta \right) \gamma ^{\dagger
}\gamma  \label{gamma2}
\end{eqnarray}
\begin{eqnarray}
i\frac{d\gamma ^{\dagger 2}}{dt} &=&-E_{1}-6g\nu e^{\theta }\sinh 2\theta
\gamma -2\left[ \Lambda +g\nu e^{\theta }\left( 2\cosh 2\theta +\sinh
2\theta \right) \right] \gamma ^{\dagger }  \nonumber \\
&&-3g\sinh ^{2}2\theta \gamma ^{2}-2\left[ E_{0}+g\left( \cosh ^{2}2\theta +%
\frac{1}{2}\sinh ^{2}2\theta \right) \right] \gamma ^{\dagger 2}  \nonumber
\\
&&-2\left( E_{1}+3g\cosh 2\theta \sinh 2\theta \right) \gamma ^{\dagger
}\gamma  \label{gamma2+}
\end{eqnarray}
\begin{equation}
i\frac{d\gamma ^{\dagger }\gamma }{dt}=\Lambda \left( \gamma ^{\dagger
}-\gamma \right) +E_{1}\left( \gamma ^{\dagger 2}-\gamma ^{2}\right)
\label{enne}
\end{equation}
Note that the truncation is performed after normal ordering of higher order
terms. Then, the equations for the bilinear operators are not equivalent to
those of the linear ones, i.e., for example, $\gamma ^{2}$ is not simply the
square of $\gamma $.

The natural choice for the parameters $\nu $ and $\theta $ cancels all the
constants appearing in the above equations which should give rise to
instability also for $\lambda \neq 0$. This choice corresponds to the constraints $%
\Lambda =0$ and $E_{1}=0$. Due to these conditions
\begin{equation}
E_{0}=\left( \frac{\lambda }{\nu }-2g\sinh 2\theta \right) e^{-2\theta }.
\label{ezero}
\end{equation}
An important consequence of these constraints is the conservation of the quasi-particles number $n_\gamma=\gamma ^{\dagger }\gamma $: 

\begin {equation}
\frac{dn_\gamma}{dt}=0
\end{equation}

 Furthermore, the value
of $\theta $ is exactly that of Eq. (\ref{theta}).

The important feature of the previous approximation is that the
continuity equation is valid independently on the satisfaction of  $%
\Lambda =0$ and $E_{1}=0$. Once the constraints are taken into account, the continuity equation implies
\begin{equation}
i\frac{\sinh
2\theta }{2}\frac{d}{dt}\left( \gamma ^{2}+\gamma ^{\dagger 2}\right)  +i\nu
e^{\theta }\frac{d}{dt}\left( \gamma +\gamma ^{\dagger }\right) =\lambda
e^{-\theta }\left( \gamma -\gamma ^{\dagger }\right) .
\end{equation}

The proof requires some tedious algebra which will be reported in appendix.
It is worth to note that for $\lambda=0$ a new constant of motion arises which is given by $(\sinh2\theta/2)\left( \gamma ^{2}+\gamma ^{\dagger 2}\right)  +\nu
e^{\theta }\left( \gamma +\gamma ^{\dagger }\right)$. 

In the Laplace space, the equations of motion can be
summarized as follows:
\begin{equation}
\omega \Gamma _{i}\left( \omega \right) =-\sum_{k=1}^{4}\phi _{ik}\Gamma
_{k}-i\Gamma _{i}\left( t=0\right)
\end{equation}
where $\Gamma _{1}=\gamma $, $\Gamma _{2}=\gamma ^{\dagger }$, $\Gamma
_{3}=\gamma ^{2}$, and $\Gamma _{4}=\gamma ^{\dagger 2}$, while the
continuity equation implies
\begin{eqnarray}
\frac{\sinh 2\theta }{2\nu }e^{-\theta }\left( \phi _{31}+\phi _{41}\right)
+\phi _{11}+\phi _{21} &=&\frac{\lambda }{\nu }e^{-2\theta } \\
\frac{\sinh 2\theta }{2\nu }e^{-\theta }\left( \phi _{32}+\phi _{42}\right)
+\phi _{12}+\phi _{22} &=&-\frac{\lambda }{\nu }e^{-2\theta } \\
+\phi _{13}+\phi _{23} &=&0 \\
\frac{\sinh 2\theta }{2\nu }e^{-\theta }\left( \phi _{34}+\phi _{44}\right)
+\phi _{14}+\phi _{24} &=&0
\end{eqnarray}
From these equations we see that it is possible to substitute the first row
of the determinant\ of coefficients $\phi _{ik}$ with the $\left\{ \left(
\lambda /\nu \right) e^{-2\theta },-\left( \lambda /\nu \right) e^{-2\theta
},0,0\right\} $. The evolution matrix in the $\omega $-space reads
\begin{equation}
D\left( \omega \right) =\left( \omega +\frac{\lambda }{\nu }e^{-2\theta
}\right) D_{11}\left( \omega \right) -\frac{\lambda }{\nu }e^{-2\theta
}D_{12}\left( \omega \right)
\end{equation}
where $D_{ik}$ is the minor associated to the matrix element $\omega \delta
_{ik}+\phi _{ik}$. It is immediately verified the existence of a pole in the
origin ($\omega =0$) for $\lambda \rightarrow 0$. This
result is expected on the basis of general arguments (Goldstone theorem).
It is worth to note that $D_{ik}\left( \omega \right) $ are finite in the
present approximation.

The resiliency of PD is directly related with the presence of the pole in
the origin. Indeed, by considering the evolution of $p\left( t\right) $ we
find
\begin{equation}
p\left( t\right) =\left( f_{1}x+f_{2}x^{2}+f_{3}p^{2}\right)
t+f_{4}p+f_{5}\left( xp+px\right)
\end{equation}
where $f_{i}$ are some nonvanishing functions of $g$, $\nu $, and $\theta $.
The presence of the term proportional to $t$ is responsible for PD. Due to
this term, the variance of $p$ on a coherent state $\left| \alpha
\right\rangle $ grows as $t^{2}$, as in the Bogoliubov case.

\section{Conclusions\label{IV}}

In this paper we discussed the problem of phase diffusion in finite-size
systems which exhibit Bose-Einstein condensation. We revised the Bogoliubov and the
Hartree-Fock-Bogoliubov approximations. While the first one satisfies at least weakly the
constraint of the continuity equation, the latter violates this
constraint. Then, we extended the self-consistent approach introducing a new approximation based on the method of equations of motion which satisfies the continuity equation. 
This approximation is developed in the space of Bogoliubov quasi-particles. The merit of the present method is to reconcile the existence of a gapless spectrum with the absence of divergencies in the vanishing symmetry-breaking limit. At the present, while the extension to systems with many degrees of freedom seems to be straightforward, the validity of PD in higher order truncation schemes is under investigation.

\acknowledgments
We are indebted with S. Paganelli for helpful discussions and comments.

\renewcommand{\theequation}{A-\arabic{equation}}
\setcounter{equation}{0} 

\section*{APPENDIX \label{A}}

In this appendix we show that the continuity equation is satisfied within
the approximation performed in this paper. The two independent conditions to be
fulfilled are
\begin{eqnarray}
\frac{\sinh 2\theta }{2\nu }e^{-\theta }\left( \phi _{32}+\phi _{42}\right)
+\phi _{12}+\phi _{22} &=&-\frac{\lambda }{\nu }e^{-2\theta }, \\
\frac{\sinh 2\theta }{2\nu }e^{-\theta }\left( \phi _{33}+\phi _{43}\right)
+\phi _{13}+\phi _{23} &=&0.
\end{eqnarray}
By writing explicitly the coefficients
\begin{equation}
\frac{\sinh 2\theta }{2\nu }e^{-\theta }\left[ 2g\nu e^{\theta }\left(
2\cosh 2\theta +\sinh 2\theta \right) -6g\nu e^{\theta }\sinh 2\theta \right]
+E_{0}=\frac{\lambda }{\nu }e^{-2\theta },
\end{equation}
\begin{eqnarray}
\frac{\sinh 2\theta }{2\nu }e^{-\theta }\left\{ 2\left[ E_{0}+g\left( \cosh
^{2}2\theta +\frac{1}{2}\sinh ^{2}2\theta \right) \right] -3g\sinh
^{2}2\theta \right\} \nonumber\\
+g\nu e^{\theta }\left( 2\cosh 2\theta +\sinh 2\theta \right) -3g\nu
e^{\theta }\sinh 2\theta &=&0.
\end{eqnarray}

The first one reduces simply to
\begin{equation}
2g\sinh 2\theta e^{-2\theta }+E_{0}=\frac{\lambda }{\nu }e^{-2\theta },
\end{equation}
which gives exactly the value of $E_{0}$ written in Eq. (\ref{ezero}) . The
second one reads
\begin{equation}
\sinh 2\theta \left( E_{0}+g\right) +2g\nu ^{2}=0.  \label{cont2}
\end{equation}
To show that this equality holds we use the condition $\Lambda =0$ putting
it in $E_{1}=0$, which lead to
\begin{equation}
2g\nu ^{2}+g\sinh 2\theta e^{-4\theta }+\frac{\lambda }{\nu }\sinh 2\theta
e^{-2\theta }=0.
\end{equation}
Due to this condition, Eq. (\ref{cont2}) becomes
\begin{equation}
\sinh 2\theta \left( \frac{\lambda }{\nu }e^{-2\theta }-2g\sinh 2\theta
e^{-2\theta }+g\right) -\frac{\lambda }{\nu }\sinh 2\theta e^{-2\theta
}-g\sinh 2\theta e^{-4\theta }=0.
\end{equation}
Now, it is simple to show that the left hand side vanishes.

\end{document}